\documentclass[amsmath,amssymb,aps,prd,eqsecnum,reprint]{revtex4-2}

\usepackage{array}
\usepackage[dvipsnames]{xcolor}
\usepackage{graphicx}
\usepackage{hyperref}
\usepackage{amsthm}

\newtheorem*{remark}{Remark}
\newtheorem{theorem}{Proposition}
\newtheorem{lemma}[theorem]{Lemma}
\newtheorem{corollary}[theorem]{Corollary}
\newtheorem{definition}[theorem]{Definition}

\newcommand{\Yslm}[3]{{}_{#1}Y_{#2}^{#3}}
\newcommand{\swsh}{SWSH }
\newcommand{\edth}[1]{{}_{#1}\eth \,}
\newcommand{\edthbar}[1]{{}_{#1}{\overline {\eth }}\,}

\newcommand{\dphi}{\left(\phi - \phi ' \right)} 

\allowdisplaybreaks

\begin{document}

\title{Some addition theorems for spin-weighted spherical harmonics}

\author{Alessandro Monteverdi}
\email{AMonteverdi1@sheffield.ac.uk}

\affiliation{School of Mathematical and Physical Sciences,
The University of Sheffield,\\
Hicks Building,
Hounsfield Road,
Sheffield. S3 7RH United Kingdom}

\author{Elizabeth Winstanley}
\email{E.Winstanley@sheffield.ac.uk}

\affiliation{School of Mathematical and Physical Sciences,
The University of Sheffield,\\
Hicks Building,
Hounsfield Road,
Sheffield. S3 7RH United Kingdom}

\date{\today}

\begin{abstract}
We present some addition theorems for spin-weighted spherical harmonics, generalizing previous results for scalar (spin-zero) spherical harmonics. 
These addition theorems involve sums over the azimuthal quantum number of products of two spin-weighted spherical harmonics at different points on the two-sphere, either (or both) of which are differentiated with respect to one of their arguments. 
\end{abstract}

\maketitle

\section{Introduction}
\label{sec:intro}

Spin-weighted spherical harmonics (SWSH) \cite{Newman:1966ub,Goldberg:1966uu} are a class of functions on the two-sphere, generalizing the  
standard (spin-$0$) spherical harmonics \cite{Muller:1966}.
While originally introduced in the context of studies of the BMS group \cite{Newman:1966ub,Goldberg:1966uu}, 
\swsh have found application in a wide range of fields \cite{TorresDelCastillo:2007}, including 
the study of anisotropies in the cosmic microwave background  \cite{Hu:1997hp,Lewis:2001hp,Wiaux:2005fp,Wiaux:2005fm,Shiraishi:2012bh,Seibert:2018};
gravitational physics, including gravitational waves and perturbations of black hole space-times \cite{Thorne:1980ru,Sharma:2011fk,GonzalezLedesma:2020dgx,Spiers:2023mor};
geosciences \cite{Michel:2020,Freeden:2022};
anisotropic turbulence \cite{Rubinstein:2015}; and
electromagnetism \cite{Scanio:1977}.
Further details of the properties of \swsh and a more extensive list of references can be found in, for example, \cite{Michel:2020,Freeden:2022}, see also \cite{Campbell:1971,Breuer:1977,Dray:1984gy,Dray:1986,Boyle:2016tjj,Lee:2023jfi} for more details.

In this paper we present some generalizations of the well-known addition theorem for \swsh \cite{Hu:1997hp,Sharma:2011fk,Michel:2020,Freeden:2022}, which we have been unable to find explicitly in the literature to date. 
It should be emphasized that our focus in this paper is \swsh functions, rather than scalar, tensor or spinor harmonics or their generalizations, for which some addition theorems
can be found in \cite{Bouzas:2011ug,Bouzas:2011uh}.

The outline of this work is as follows. 
In Sec.~\ref{sec:physics} we discuss some of the applications of \swsh in physics and motivate the addition theorems which are the focus of our work.
Sec.~\ref{sec:swsh} reviews the key definitions and properties of \swsh which are required for our analysis. 
For ease of reference, our new addition theorems are presented in Sec.~\ref{sec:results}, prior to their derivation in Sec.~\ref{sec:proofs}.
Our conclusions can be found in Sec.~\ref{sec:conc}.

\section{Physical applications of \swsh addition theorems}
\label{sec:physics}

\swsh typically arise in physical applications when one is considering a quantity (such as a field perturbation) which depends on spherical polar coordinates. 
Let us give a few examples:
\begin{description}
    \item[Data analysis]
    Data from surveys such as those of the cosmic microwave background \cite{Lewis:2001hp} or satellite gravity gradiometry \cite{Seibert:2018} is naturally defined on a two-sphere.  
    An expansion of observables in terms of \swsh facilitates computationally-efficient methods of analyzing such data \cite{Wiaux:2005fm,Wiaux:2005fp}. 
    \item[Modelling anisotropies]
    When physical systems are not spherically symmetric, an expansion in \swsh provides a powerful framework for the study of anisotropies, for example in fluid turbulence \cite{Rubinstein:2015} or the cosmic microwave background \cite{Hu:1997hp}, since the coefficients of the \swsh in such an expansion correspond to excitations of a particular multipole \cite{Scanio:1977,Thorne:1980ru}. 
    \item[Perturbations of black holes]
    Classical field equations on black hole backgrounds play an important role many aspects of general relativity, including the modelling of gravitational waves.
    If the background black hole is spherically symmetric, then separable mode solutions of the classical field equation involve \swsh  (see, for example, \cite{Spiers:2023mor}). 
\end{description}

Our focus in this paper is addition theorems for {\mbox {\swsh\!.}} 
The theorems of the type we consider involve a product of two \swsh at different points over the two-sphere, summed over one of the quantum numbers of the \swsh (see Sec.~\ref{sec:swsh} for more precise details). 
Such products of two \swsh at different points arise, for example, when one is interested in the Green function for a classical field equation on a flat or curved space-time.
A Green function is a distributional solution of a classical field equation which depends on two points in space-time. 
In applications, it is often useful to have the Green function written as a sum over separable mode solutions of the field equation, see, for example, \cite{Thorne:1980ru}.
When the mode solutions involve \swsh\!, the Green function therefore involves a product of \swsh at different points on the two-sphere, and the quantum numbers in the \swsh label the field modes and are summed over. 
Addition theorems of the type considered in this paper are particularly useful in situations where the background on which the field propagates (such as a black hole space-time) is spherically symmetric, in which case the functions other than the \swsh in the field modes do not depend on one of the quantum numbers (in particular, the azimuthal quantum number $m$ as defined in Sec.~\ref{sec:swsh}) in the \swsh\!.
In this case, an addition theorem involving a sum over $m$ enables the Green function to be simplified.

We close this section by giving more details of an application of our addition theorems for \swsh in quantum field theory in curved space-time \cite{Balakumar:2022yvx,Alvarez,Monteverdi}, in particular the computation of expectation values of observables such as the stress-energy tensor.
The stress-energy tensor expectation value is of central importance in quantum field theory on curved space-time, since it governs, via the semiclassical Einstein equations, the backreaction of the quantum field on the curved space-time.
To find the stress-energy tensor expectation value, one starts with the Green function for the classical equation of the particular field under consideration (which, as explained above, involves products of \swsh at different points on the two-sphere).
One then applies a second-order differential operator to the Green function, with the result that one or more derivatives act on the \swsh in the sum. 
If one considers a field of nonzero spin on a four-dimensional black hole (see, for example, \cite{Alvarez}), the resulting expectation values involve sums over \swsh of precisely the form we prove here (see Corollary \ref{QFTresults} in Sec.~\ref{sec:results} below) for a particular value of the spin-weight, while a scalar field on a particular five-dimensional black hole \cite{Monteverdi} has a tower of modes involving \swsh of arbitrary (integer or half-integer) spin, with the result that we require the addition theorems in Corollary \ref{QFTresults} for general spin-weight.

Having motivated the \swsh addition theorems which we will derive in this paper, in the next section we review the salient properties of \swsh\!, before turning to the addition theorems themselves.

\section{Properties of \swsh}
\label{sec:swsh}

The \swsh $\Yslm{s}{\ell }{m}(\theta , \phi )$ are functions on the two-sphere ${\mathbb{S}}^{2}$ which have spin-weight $s$, where $s$ is an integer or half-integer \cite{Newman:1966ub,Goldberg:1966uu}.
We employ the usual spherical polar coordinates $(\theta , \phi )$ on ${\mathbb{S}}^{2}$, with $\theta \in [0,\pi]$ and $\phi \in [0,2\pi )$.
The polar angle $\theta $ is the angle between the vector joining the point on the surface of ${\mathbb{S}}^{2}$ to the origin and the $z$-axis; while $\phi $ is the azimuthal angle in the $(x,y)$-plane, so that a point on ${\mathbb{S}}^{2}$ has Cartesian coordinates $(\sin \theta \cos \phi, \sin \theta \sin \phi, \cos \theta )$.

The \swsh depend on three quantum numbers $s$, $\ell $ and $m$.
The spin $s$ is a positive or negative integer or half-integer.  
The orbital angular momentum quantum number $\ell $ then takes values $\ell = |s|, |s|+1, |s|+2, \ldots $, and the azimuthal quantum number $m$ takes the values $m=-\ell, -\ell + 1, \ldots \ell - 1, \ell $.  

There are a number of different definitions of the \swsh in the literature. In this work, we follow the conventions of \cite{Michel:2020}, in particular:
\begin{definition}
\label{def:swsh}
The \swsh $\Yslm{s}{\ell }{m}(\theta , \phi )$ are defined in terms of the Wigner $D$-matrices by \cite{Michel:2020}
\begin{equation}
    \Yslm{s}{\ell }{m}(\theta , \phi )= \left( -1 \right) ^{s} {\sqrt {\frac{2\ell + 1}{4\pi }}}
    D^{\ell *}_{m, -s}(\phi , \theta, 0),
    \label{eq:swshdef}
\end{equation}
where $*$ denotes complex conjugation.
The Wigner $D$-matrices take the form \cite{Michel:2020}
\begin{equation}
    D^{\ell *}_{m, -s}(\phi , \theta, \chi ) = 
    {\rm {e}}^{{\rm {i}}m\phi  } d_{m,-s}^{\ell }(\theta ){\rm {e}}^{-{\rm {i}}s\chi  }
\end{equation}
where $d_{m,-s}^{\ell }(\theta )$ are real functions given by \cite{Michel:2020}
\begin{multline}
    d_{m,-s}^{\ell }(\theta ) =
     \frac{(-1)^{\ell + m}}{2^{\ell }}
     {\sqrt {\frac{\left( \ell - m \right) !}{\left( \ell + m \right)! \left( \ell + s \right)! \left(  \ell - s \right) !}}}
     \\ \times 
     \left( 1 - \cos \theta \right) ^{\frac{1}{2}\left( m+s\right) } \left( 1+\cos \theta \right) ^{\frac{1}{2}\left( m-s \right) }
     \\
     \times
     \left(  \frac{d}{d\left[ \cos \theta \right] }\right) ^{\ell + m }
     \left[ 
     \left(  1 - \cos \theta \right) ^{\ell - s}
     \left( 1 + \cos \theta \right) ^{\ell + s}
     \right] .
     \label{eq:WignerD}
\end{multline}
\end{definition}

\begin{remark}
Our results in this paper are valid for both integer and half-integer spins $s$. When $s$ is a half-integer, the factor of $(-1)^{s}$ in \eqref{eq:swshdef} will be purely imaginary (see, for example, \cite{Lee:2023jfi} for some specific examples of \swsh in this case).
The quantities $\ell \pm m$, $\ell \pm s$ and $m \pm s$ appearing in the Wigner $D$-matrices \eqref{eq:WignerD} are always integers.
\end{remark}

Our derivation of new addition theorems rests on the action of the operators $\edth{s}$ and $\edthbar{s}$, as defined below. 

\begin{definition}
	\label{def:edth}
The operators $\edth{s}$ and $\edthbar{s}$ act on the \swsh as follows \cite{Newman:1966ub,Goldberg:1966uu,Michel:2020}:
\begin{subequations}
\label{eq:edthdef}
\begin{align}
    \edth{s} \Yslm{s}{\ell}{m} (\theta, \phi )   & = - \left[ \partial _{\theta } + \frac{{\rm {i}}}{\sin \theta } \partial _{\phi } - s \cot \theta  \right]  \Yslm{s}{\ell}{m} (\theta, \phi ) ,
    \label{eq:edth}
    \\
    \edthbar{s} \Yslm{s}{\ell}{m} (\theta, \phi )   & = - \left[ \partial _{\theta } - \frac{{\rm {i}}}{\sin \theta } \partial _{\phi } + s \cot \theta  \right]  \Yslm{s}{\ell}{m} (\theta, \phi ) .
    \label{eq:edthbar}
\end{align}
\end{subequations}
\end{definition}

\begin{remark}
	To avoid confusion, we denote by $\edth{s'}'$, $\edthbar{s'}'$ the operators $\edth{s} $ and $\edthbar{s}$ with $(\theta , \phi )$ replaced by $(\theta' , \phi ')$ and with spin $s'$ instead of $s$.
\end{remark}

\begin{theorem}
\label{thm:raiselower}
The operators $\edth{s}$, $\edthbar{s}$ respectively increase and decrease the spin-weight $s$ of an \swsh $\Yslm{s}{\ell}{m}$ by one \cite{Newman:1966ub,Goldberg:1966uu,Michel:2020,Freeden:2022}:
\begin{subequations}
\label{eq:raiselower}
    \begin{align}
        \edth{s} \Yslm{s}{\ell}{m} & = \Yslm{s+1}{\ell}{m} {\sqrt {(\ell - s)(\ell + s+ 1)}}  ,
        \label{eq:raiselower1}
        \\
        \edthbar{s} \Yslm{s}{\ell}{m} & = -\Yslm{s-1}{\ell}{m} {\sqrt {(\ell + s)(\ell - s+ 1)}} .
        \label{eq:raiselower2}
    \end{align}
\end{subequations} 
\end{theorem}

The following Lemma (which follows straightforwardly from Definition \ref{def:edth}) will be central to the derivation of our addition theorems.

\begin{lemma}
	\begin{subequations}
		\begin{align}
			\frac{\partial }{\partial \theta } \Yslm{s}{\ell }{m}(\theta , \phi ) & = 
			-\frac{1}{2} \left[ \edth{s} + \edthbar{s} \right] \Yslm{s}{\ell }{m}(\theta , \phi ),
			\\
			\frac{\partial }{\partial \phi }  \Yslm{s}{\ell }{m}(\theta , \phi ) & = 
			\frac{{\rm {i}}}{2} \sin \theta \left[ \edth{s} - \edthbar{s} \right] \Yslm{s}{\ell }{m}(\theta , \phi )
            \nonumber \\ &  \qquad 
			- {\rm {i}}s \cos \theta \Yslm{s}{\ell }{m}(\theta , \phi )  .
		\end{align}
	\end{subequations}
\end{lemma}

The well-known usual addition theorem \cite{Hu:1997hp,Sharma:2011fk,Michel:2020,Freeden:2022} for \swsh depends on the Euler angles $\alpha $, $\beta $, $\gamma$, defined below.

\begin{definition}
	\label{def:euler}
	The Euler angles $\alpha $, $\beta $, $\gamma $ are given in terms of angles $(\theta, \phi )$ and $(\theta' , \phi ')$ as follows \cite{Michel:2020}:	
	\begin{subequations}
		\label{eq:euler}
		\begin{align}
			\cot \alpha  & = \cos \theta  \cot \dphi  
			- \cot \theta '\sin \theta  \csc \dphi ,
			\label{eq:alpha}
			\\
			\cos \beta  & = \cos \theta \cos \theta ' + \sin \theta \sin \theta ' \cos \dphi ,
			\label{eq:beta}
			\\
			\cot \gamma & = \cos \theta '\cot \dphi - \cot \theta  \sin \theta ' \csc \dphi .
			\label{eq:gamma}
		\end{align}
	\end{subequations}
\end{definition}

\begin{remark}
    The definitions \eqref{eq:euler} are, strictly speaking, valid only when $\phi - \phi '$ is not a multiple of $\pi $.
In the coincidence limit $\theta ' \rightarrow \theta $, $\phi ' \rightarrow \phi $, taking the appropriate limit of \eqref{eq:euler} gives $\alpha = \beta =\gamma = 0$.
\end{remark}

We are now in a position to state the usual addition theorem for \swsh\!\!.

\begin{theorem}
\label{thm:addition}
	The addition theorem for spin-weighted spherical harmonics is \cite{Michel:2020}:
\begin{multline}
\left( -1 \right)^{s} \sum _{m=-\ell }^{\ell } \Yslm{s}{\ell }{m} (\theta , \phi ) \Yslm{s'}{\ell }{m*}(\theta' , \phi ') 
\\
= {\sqrt {\frac{2\ell + 1}{4\pi }}} {\rm {e}}^{-{\rm {i}}s\alpha } \Yslm{s}{\ell }{-s'} (\beta, \gamma ) .
\label{eq:swshaddition}
\end{multline}
\end{theorem}

\begin{remark}
	The addition theorem \eqref{eq:swshaddition} is valid only when the spins $s$, $s'$ differ by an integer, and $\ell \ge \max \{ |s|, |s'| \}$.	
 We have written \eqref{eq:swshaddition} in an alternative, but equivalent, form to that displayed in \cite{Michel:2020}. 
 The proof of Theorem \ref{thm:addition} in \cite{Michel:2020} is presented only for integer spins, but the result is equally valid for half-integer spins and can be straightforwardedly derived using, for example, Definition \ref{def:swsh} and the properties of the Wigner $D$-functions \cite{Michel:2020,Khersonskii:1988krb}.
 The addition theorem  takes a slightly different form in some references \cite{Hu:1997hp,Sharma:2011fk} due to differences in the definition of the \swsh \!\!.
\end{remark}

To take the coincidence limit $\theta ' = \theta $, $\phi ' = \phi $ in the addition theorem, we require the coincidence limit of the \swsh\!\!:

\begin{lemma}
We have \cite{Michel:2020}
\begin{align}
    \label{eq:swsh2}
    \Yslm{s}{\ell }{m}(0,0) & = \left( -1 \right) ^{s} \delta _{s,-m}{\sqrt {\frac{2\ell + 1}{4\pi }}},
    \\
    \nonumber
\end{align}
where $\delta _{s,s'}$ is the usual Kronecker delta:
\begin{equation}
	\delta _{s,s'} = \begin{cases}
		1 & {\mbox {if $s'=s$}}, \\
		0 & {\mbox {otherwise}}.
	\end{cases}
 \end{equation}
 \end{lemma}

\begin{corollary}
\label{thm:corollary}
The coincidence limit $\theta ' = \theta $, $\phi ' = \phi $ of the addition theorem \eqref{eq:swshaddition} is:
\begin{equation}
    \sum _{m=-\ell }^{\ell } \Yslm{s}{\ell }{m} (\theta , \phi ) \Yslm{s'}{\ell }{m*}(\theta , \phi )  
= \frac{2\ell + 1}{4\pi } \delta _{s,s'} .	
\end{equation}	
\end{corollary}

Our purpose in this note is to derive generalizations of Proposition \ref{thm:addition} and Corollary \ref{thm:corollary} which involve a derivative of either (or both) of the \swsh in the sum.

\bigskip
 
\section{Further addition theorems}
\label{sec:results}

We now summarize our results, which will be derived in the following section. 
To the best of our knowledge, these have not previously appeared explicitly in the literature.
These results are valid under the same conditions on the quantum numbers $s$, $s'$ and $\ell $ as for the original addition theorem, namely the spins $s$, $s'$ can only differ by an integer, and it must be the case that $\ell \ge \max \{ |s|, |s'| \}$.

\begin{widetext}

\begin{theorem}
	\label{results}
	\begin{subequations}
	\label{eq:results}
	\begin{align}
		 & (-1)^{s} \sum _{m=-\ell }^{\ell } \left[ \frac{\partial }{\partial \theta }\Yslm{s}{\ell }{m} (\theta , \phi ) \right] 
		\Yslm{s'}{\ell }{m*}(\theta ', \phi ')		
 \nonumber \\ &  = 
		\frac{1}{2}{\sqrt {\frac{2\ell + 1}{4\pi }}}  \left\{   
		{\sqrt {\left( \ell -s \right)\left( \ell + s + 1 \right)}}\, {\rm {e}}^{-{\rm {i}}(s+1)\alpha } \Yslm{s+1}{\ell }{-s'} (\beta, \gamma ) 
		-{\sqrt {\left( \ell +s \right)\left( \ell - s + 1 \right)}} \, {\rm {e}}^{-{\rm {i}}(s-1)\alpha } \Yslm{s-1}{\ell }{-s'} (\beta, \gamma )
		\right\} ,
		\label{eq:addition1}
  \\ \nonumber 
		\\
			 & (-1)^{s} \sum _{m=-\ell }^{\ell } m \, \Yslm{s}{\ell }{m} (\theta , \phi ) \Yslm{s'}{\ell }{m*}(\theta ', \phi ')
  \nonumber \\ & = 
			-\frac{1}{2}{\sqrt {\frac{2\ell + 1}{4\pi }}}  \left\{   
			{\sqrt {\left( \ell -s \right)\left( \ell + s + 1 \right)}} \, {\rm {e}}^{-{\rm {i}}(s+1)\alpha } \Yslm{s+1}{\ell }{-s'} (\beta, \gamma )  \sin \theta 
   \right. \nonumber \\ &  \qquad  \qquad \qquad \left.
			+{\sqrt {\left( \ell +s \right)\left( \ell - s + 1 \right)}} \, {\rm {e}}^{-{\rm {i}}(s-1)\alpha } \Yslm{s-1}{\ell }{-s'} (\beta, \gamma )
			 \sin \theta   
			 +2 s \, {\rm {e}}^{-{\rm {i}}s\alpha } \Yslm{s}{\ell }{-s'} (\beta, \gamma )\cos \theta 
			\right\} ,
		\label{eq:addition2}
   \\ \nonumber 
		\\
			& (-1)^{s}\sum _{m=-\ell }^{\ell } \left[ \frac{\partial }{\partial \theta }\Yslm{s}{\ell }{m} (\theta , \phi ) \right] 
		\left[ \frac{\partial }{\partial \theta '}	\Yslm{s'}{\ell }{m*}(\theta ', \phi ')  \right]
		\nonumber \\ &  = 
		 -\frac{1}{4} {\sqrt {\frac{2\ell + 1}{4\pi }}}  
		 \left[
		{\sqrt {\left(\ell - s\right)\left( \ell + s+ 1 \right) \left(\ell - s'\right)\left( \ell + s'+ 1 \right) }} \, 
		{\rm {e}}^{-{\rm {i}}(s+1)\alpha }\Yslm{s+1}{\ell }{-s'-1}(\beta,\gamma )
		\right. \nonumber \\ & \left.  \qquad 
		-{\sqrt {\left(\ell - s\right)\left( \ell + s+ 1 \right) \left(\ell + s'\right)\left( \ell - s'+ 1 \right) }} \, 
		{\rm {e}}^{-{\rm {i}}(s+1)\alpha }\Yslm{s+1}{\ell }{-s'+1}(\beta,\gamma )
		\right. \nonumber \\ & \left.  \qquad
		-{\sqrt {\left(\ell + s\right)\left( \ell - s+ 1 \right) \left(\ell - s'\right)\left( \ell + s'+ 1 \right) }} \, 
		{\rm {e}}^{-{\rm {i}}(s-1)\alpha }\Yslm{s-1}{\ell }{-s'-1}(\beta,\gamma )
		\right. \nonumber \\ &  \left.  \qquad 
		+{\sqrt {\left(\ell + s\right)\left( \ell - s+ 1 \right) \left(\ell + s'\right)\left( \ell -s'+ 1 \right) }} \, 
		{\rm {e}}^{-{\rm {i}}(s-1)\alpha }\Yslm{s-1}{\ell }{-s'+1}(\beta,\gamma )
		\right] ,
		\label{eq:addition3}
   \\ \nonumber 
		\\
				 & (-1)^{s}\sum _{m=-\ell }^{\ell } m^{2} \Yslm{s}{\ell }{m} (\theta , \phi )\Yslm{s'}{\ell }{m*}(\theta ', \phi ')
		\nonumber \\ &  =
		-{\sqrt {\frac {2\ell +1}{4\pi }}}\bigg\{ 
		\frac{1}{4}\sin \theta \sin \theta ' \left[ 
		{\sqrt {\left(\ell - s\right)\left( \ell + s+ 1 \right) \left(\ell - s'\right)\left( \ell + s'+ 1 \right) }} \,
		{\rm {e}}^{-{\rm {i}}(s+1)\alpha } \Yslm{s+1}{\ell }{-s'-1}(\beta  , \gamma )
		\right. \nonumber \\ & \left.  \qquad  
		+{\sqrt {\left(\ell - s\right)\left( \ell + s+ 1 \right) \left(\ell + s'\right)\left( \ell - s'+ 1 \right) }} \, 
		{\rm {e}}^{-{\rm {i}}(s+1)\alpha } \Yslm{s+1}{\ell }{-s'+1}(\beta  , \gamma )
		\right. \nonumber \\ & \left.   \qquad  
		+{\sqrt {\left(\ell + s\right)\left( \ell - s+ 1 \right) \left(\ell - s'\right)\left( \ell + s'+ 1 \right) }} \, 
		{\rm {e}}^{-{\rm {i}}(s-1)\alpha }\Yslm{s-1}{\ell }{-s'-1}(\beta  , \gamma )
		\right. \nonumber \\ & \left.  \qquad 
		+{\sqrt {\left(\ell + s\right)\left( \ell - s+ 1 \right) \left(\ell + s'\right)\left( \ell -s'+ 1 \right) }} \, 
		{\rm {e}}^{-{\rm {i}}(s-1)\alpha }\Yslm{s-1}{\ell }{-s'+1}(\beta  , \gamma )
		\right]
		\nonumber \\ & \qquad   
		- \frac{s'}{2} \sin \theta \cos \theta ' \left[ 
		{\sqrt {\left(\ell - s\right)\left( \ell + s+ 1 \right) }} \, 
		{\rm {e}}^{-{\rm {i}}(s+1)\alpha }\Yslm{s+1}{\ell }{-s'}(\beta  , \gamma )
		+ {\sqrt {\left(\ell + s\right)\left( \ell - s+ 1 \right) }} \, 
		{\rm {e}}^{-{\rm {i}}(s-1)\alpha }\Yslm{s-1}{\ell }{-s'}(\beta  , \gamma ) 
		\right] 
		\nonumber \\ & \qquad 
		+ \frac{s}{2} \cos \theta \sin \theta ' \left[
		{\sqrt {\left(\ell - s'\right)\left( \ell + s'+ 1 \right) }} \, 
		{\rm {e}}^{-{\rm {i}}s\alpha }\Yslm{s}{\ell }{-s'-1}(\beta  , \gamma ) 
		+ {\sqrt {\left(\ell + s'\right)\left( \ell - s'+ 1 \right) }} \, 
		{\rm {e}}^{-{\rm {i}}s\alpha }\Yslm{s}{\ell }{-s'+1}(\beta  , \gamma ) 
		\right] 
		\nonumber \\ & \qquad
		- ss' {\rm {e}}^{-{\rm {i}}s\alpha }\Yslm{s}{\ell }{-s'}(\beta , \gamma ) \cos \theta \cos \theta '
		\bigg\} ,
		\label{eq:addition4}
   \\ \nonumber 
		\\
		 & (-1)^{s}\sum _{m=-\ell }^{\ell } m\left[ \frac{\partial }{\partial \theta }\Yslm{s}{\ell }{m} (\theta , \phi ) \right] 
		\Yslm{s'}{\ell }{m*}(\theta ', \phi ') &
		\nonumber \\ &  =
		-\frac{1}{4} {\sqrt {\frac {2\ell +1}{4\pi }}} \bigg\{ 
		\left[
		{\sqrt {\left(\ell - s\right)\left( \ell + s+ 1 \right) \left(\ell - s'\right)\left( \ell + s'+ 1 \right) }} \, 
		{\rm {e}}^{-{\rm {i}}(s+1)\alpha }\Yslm{s+1}{\ell }{-s'-1}(\beta , \gamma ) 
		\right. \nonumber \\ & \left. \qquad
		- {\sqrt {\left(\ell + s\right)\left( \ell - s+ 1 \right) \left(\ell - s'\right)\left( \ell + s'+ 1 \right) }} \,
		{\rm {e}}^{-{\rm {i}}(s-1)\alpha }\Yslm{s-1}{\ell }{-s'-1}(\beta , \gamma ) 
		\right. \nonumber \\ & \left. \qquad
		+ {\sqrt {\left(\ell - s\right)\left( \ell + s+ 1 \right) \left(\ell + s'\right)\left( \ell - s'+ 1 \right) }} \,
		{\rm {e}}^{-{\rm {i}}(s+1)\alpha }\Yslm{s+1}{\ell }{-s'+1}(\beta , \gamma ) 
		\right. \nonumber \\ &  \left. \qquad
		- {\sqrt {\left(\ell + s\right)\left( \ell - s+ 1 \right) \left(\ell + s'\right)\left( \ell - s'+ 1 \right) }} \,
		{\rm {e}}^{-{\rm {i}}(s-1)\alpha }\Yslm{s-1}{\ell }{-s'+1}(\beta , \gamma ) 
		\right] \sin \theta '
		\nonumber \\ & \qquad
		-2s' \left[  {\sqrt {\left(\ell - s\right)\left( \ell + s+ 1 \right) }} \, 
		{\rm {e}}^{-{\rm {i}}(s+1)\alpha }\Yslm{s+1}{\ell }{-s'}(\beta , \gamma ) 
		- {\sqrt {\left(\ell + s\right)\left( \ell - s+ 1 \right) }} \, 
		{\rm {e}}^{-{\rm {i}}(s-1)\alpha }\Yslm{s-1}{\ell }{-s'}(\beta , \gamma ) 
		\right] \cos \theta '
		\bigg\} .
		\label{eq:addition5}
	\end{align}
	\end{subequations}
\end{theorem}

\begin{remark}
    The addition theorems \eqref{eq:results} all involve one derivative of a \swsh\!\!. 
    Similar theorems involving two or more derivatives acting on the same \swsh can be deduced from the above, together with the differential equation governing the \swsh \cite{Breuer:1977,Lewis:2001hp,Seibert:2018,Michel:2020,Freeden:2022}
    \begin{equation}   
   \left\{ \frac{1}{\sin \theta }\partial _{\theta } \left[ \sin \theta \, \partial _{\theta } \right]
  - \frac{1}{\sin ^{2} \theta } \left[ s^{2}-2{\rm {i}}s \cos \theta \,  \partial _{\phi }
   - \partial ^{2}_{\phi }\right]
   \right\} \Yslm{s}{\ell }{m}(\theta, \phi ) = \ell \left[ \ell + 1 \right] \Yslm{s}{\ell }{m}(\theta, \phi ),
    \label{eq:de}
    \end{equation}
    and noting that $\partial _{\phi }\left[ \Yslm{s}{\ell }{m}(\theta, \phi ) \right] = {\rm {i}} m\Yslm{s}{\ell }{m}(\theta, \phi ) $.
\end{remark}

Taking the coincidence limit $\theta ' = \theta $, $\phi ' = \phi $ in each of the results in \eqref{eq:results} gives, respectively, using \eqref{eq:swsh2}:
\begin{corollary}
	\label{coincidence}
	\begin{subequations}
		\label{eq:coincidence}
		\begin{align}
		\sum _{m=-\ell }^{\ell } \left[ \frac{\partial }{\partial \theta }\Yslm{s}{\ell }{m} (\theta , \phi ) \right] 
		\Yslm{s'}{\ell }{m*}(\theta , \phi )
		& = 
		-\frac{2\ell + 1}{8\pi }  \left\{   
		{\sqrt {\left( \ell -s \right)\left( \ell + s + 1 \right)}} \, \delta _{s+1,s'}
		-{\sqrt {\left( \ell -s' \right)\left( \ell + s' + 1 \right)}} \, \delta _{s,s'+1}
		\right\} ,
		\label{eq:coincidence1}	
		\\
			\sum _{m=-\ell }^{\ell } m \, \Yslm{s}{\ell }{m} (\theta , \phi ) \Yslm{s'}{\ell }{m*} (\theta , \phi )
			& = \frac{2\ell + 1}{8\pi } \left\{   
			{\sqrt {\left( \ell -s \right)\left( \ell + s + 1 \right)}} \, \delta _{s+1,s'}
			+{\sqrt {\left( \ell -s' \right)\left( \ell +s' + 1 \right)}} \, \delta _{s,s'+1}
			\right\}  \sin \theta  
			\nonumber \\ &  \qquad 
			 -\frac{2\ell + 1}{4\pi } s \,\delta _{s,s'}\cos \theta ,
			\label{eq:coincidence2}
			\\
				\sum _{m=-\ell }^{\ell } \left[ \frac{\partial }{\partial \theta }\Yslm{s}{\ell }{m} (\theta , \phi ) \right] 
			\left[ \frac{\partial }{\partial \theta }	\Yslm{s'}{\ell }{m*}(\theta , \phi )  \right]
			& = 
			\frac{2\ell + 1}{16\pi } 
			\left[
			2\left(\ell ^{2} +\ell - s^{2}\right) \, 
			\delta _{s,s'}
			\right. \nonumber \\ & \left. \qquad
			-{\sqrt {\left( \ell - s-1 \right)\left(\ell - s\right)\left( \ell + s+ 1 \right) \left(\ell + s+2\right) }} \, 
			\delta _{s+2,s'}
			\right. \nonumber \\ & \left. \qquad
			-{\sqrt {\left( \ell - s'-1 \right) \left(\ell - s'\right)\left( \ell + s'+ 1 \right) \left(\ell + s'+2\right) }} \, 
			\delta _{s,s'+2}
			\right] ,
			\label{eq:coincidence3}
			\\
				\sum _{m=-\ell }^{\ell } m^{2} \Yslm{s}{\ell }{m} (\theta , \phi )\Yslm{s'}{\ell }{m*}(\theta , \phi )
			\nonumber \\ & \hspace{-3.5cm} =
			\frac {2\ell +1}{4\pi }\bigg\{ 
		 \left[ 
				\frac{1}{2} \left(  \ell^{2}  + \ell  - s^{2} \right) \sin ^{2}\theta + s^{2}\cos ^{2} \theta  \right]
		\, \delta _{s,s'}
			\nonumber \\ & \hspace{-3cm}   
			+\frac{1}{4}\sin ^{2}\theta  \left[ {\sqrt {\left(\ell - s\right)\left( \ell + s+ 1 \right) \left(\ell + s'\right)\left( \ell - s'+ 1 \right) }} \, 
		\delta _{s+1,s'-1}
			\right. \nonumber \\ & \hspace{-2cm}    \left. 
			+{\sqrt {\left(\ell + s\right)\left( \ell - s+ 1 \right) \left(\ell - s'\right)\left( \ell + s'+ 1 \right) }} \right]  \, 
			\delta _{s-1,s'+1}
			 \nonumber \\ & \hspace{-3cm} 
			- \frac{1}{2} \sin \theta  \cos \theta  \left[
			\left( 2 s'+1 \right) {\sqrt {\left(\ell - s'\right)\left( \ell + s'+ 1 \right) }} \, 
			\delta _{s,s'+1} 
			+ \left( 2s+1 \right) {\sqrt {\left(\ell - s\right)\left( \ell +s+ 1 \right) }} \, 
			\delta _{s+1,s'}
			\right] 
			\bigg\} ,	
				\label{eq:coincidence4}
				\\ 
				\sum _{m=-\ell }^{\ell } m\left[ \frac{\partial }{\partial \theta }\Yslm{s}{\ell }{m} (\theta , \phi ) \right] 
				\Yslm{s'}{\ell }{m*}(\theta , \phi ) &
				\nonumber \\ & \hspace{-4cm} =
				\frac {2\ell +1}{16\pi } \bigg\{ 
				\left[ -2s 
				 \, \delta _{s,s'}
				- {\sqrt {\left( \ell - s'- 1 \right) \left(\ell - s'\right)\left( \ell + s'+ 1 \right)\left(\ell + s'+2\right) }} \, \delta _{s,s'+2}
				\right. \nonumber \\ & \hspace{-2cm} \left.
				+ {\sqrt {\left( \ell - s - 1 \right) \left(\ell - s\right)\left( \ell + s+ 1 \right) \left(\ell + s+2\right)}} \, \delta _{s+2,s'}
				\right] \sin \theta 
				\nonumber \\ & \hspace{-2cm} 
				-2s' \left[  {\sqrt {\left(\ell - s\right)\left( \ell + s+ 1 \right) }} \, 
				\delta _{s+1,s'}
				- {\sqrt {\left(\ell -s'\right)\left( \ell + s'+ 1 \right) }} \, 
				\delta _{s,s'+1}
				\right] \cos \theta 
				\bigg\} .
				\label{eq:coincidence5}
		\end{align}
	\end{subequations}
\end{corollary}

Finally, if we also set the spins equal, $s'=s$, the results in Corollary~\ref{coincidence} reduce further, giving, respectively:
\begin{corollary}
\label{QFTresults}
	\begin{subequations}
		\label{eq:spinsame}
		\begin{align}
			\sum _{m=-\ell }^{\ell } \left[ \frac{\partial }{\partial \theta }\Yslm{s}{\ell }{m} (\theta , \phi ) \right] 
			\Yslm{s}{\ell }{m*}(\theta , \phi ) & = 0 ,
			\label{eq:spin1}
			\\
			\sum _{m=-\ell }^{\ell } m \, \left|  \Yslm{s}{\ell }{m} (\theta , \phi ) \right| ^{2}
			& = 
			- \frac{\left( 2\ell + 1 \right) s}{4\pi }  \cos \theta ,
			\label{eq:spin2}
			\\
				\sum _{m=-\ell }^{\ell } \left| \frac{\partial }{\partial \theta }\Yslm{s}{\ell }{m} (\theta , \phi ) \right| ^{2}
				& = \frac{2\ell + 1}{8\pi } \left(\ell ^{2} +\ell - s^{2}\right) ,
				\label{eq:spin3}
				\\
				\sum _{m=-\ell }^{\ell } m^{2} \, \left|  \Yslm{s}{\ell }{m} (\theta , \phi ) \right| ^{2}
				& = \frac {2\ell +1}{8\pi } \left[ 
				\left(  \ell^{2}  + \ell  - s^{2} \right) \sin ^{2}\theta  + 2s^{2}\cos ^{2} \theta  \right] ,
				\label{eq:spin4}
				\\
			\sum _{m=-\ell }^{\ell } m\left[ \frac{\partial }{\partial \theta }\Yslm{s}{\ell }{m} (\theta , \phi ) \right] 
			\Yslm{s}{\ell }{m*}(\theta , \phi ) & = 
			-\frac {\left( 2\ell +1\right) s}{8\pi }  \sin \theta .
				\label{eq:spin5}
  		\end{align}
	\end{subequations}
\end{corollary}

When $s=0$, the results \eqref{eq:spinsame} reduce to those found in, for example, App.~C of 
\cite{Balakumar:2022yvx} (see also \cite[\S 5.10]{Khersonskii:1988krb}).

\begin{remark}
    The precise form of the addition theorems (\ref{eq:results}, \ref{eq:coincidence}) depends on the conventions used for the definition of the \swsh\!\!.
    However, the final results in \eqref{eq:spinsame} are independent of the choice of phase in the \swsh\!\!.
\end{remark}

\section{Derivation of new addition theorems}
\label{sec:proofs}

Our overall strategy in deriving the results in Proposition~\ref{results} is to apply appropriate combinations of the operators $\edth{s}$, $\edthbar{s}$, $\edth{s'}'$, and $\edthbar{s'}'$ (Definition \ref{def:edth}) to the original addition theorem \eqref{eq:swshaddition}. 
We then employ the ``raising and lowering'' properties of these operators (Proposition \ref{thm:raiselower}). 

\begin{proof}[Derivation of \eqref{eq:addition1}]
\begin{align}
	(-1)^{s}\sum _{m=-\ell }^{\ell } \left[ \frac{\partial }{\partial \theta }\Yslm{s}{\ell }{m} (\theta , \phi ) \right] 
	\Yslm{s'}{\ell }{m*}(\theta ', \phi ')
	& = 
	-\frac{(-1)^{s}}{2}  \sum _{m=-\ell }^{\ell } \left[ \edth{s}  \Yslm{s}{\ell }{m}(\theta , \phi ) + \edthbar{s}  \Yslm{s}{\ell }{m}(\theta , \phi ) \right] 	\Yslm{s'}{\ell }{m*}(\theta ', \phi ')
	\nonumber \\ & \hspace{-5cm} 
 = 
	-\frac{(-1)^{s}}{2} \sum _{m=-\ell }^{\ell } \left[ 
 {\sqrt {\left(\ell - s\right)\left( \ell + s+ 1 \right) }} \, \Yslm{s+1}{\ell }{m}(\theta ,\phi )
 		- {\sqrt {\left(\ell + s\right)\left( \ell - s+ 1 \right) }} \, \Yslm{s-1}{\ell }{m}(\theta ,\phi )
 \right] \Yslm{s'}{\ell }{m*}(\theta ', \phi ')
 	\nonumber \\ & \hspace{-5cm} 
 =  \frac{1}{2} \left\{ 
 {\sqrt {\left(\ell - s\right)\left( \ell + s+ 1 \right) }}\sum _{m=-\ell }^{\ell }  (-1)^{s+1} \Yslm{s+1}{\ell }{m}(\theta ,\phi )  \Yslm{s'}{\ell }{m*}(\theta ', \phi ')
 \right. \nonumber \\ & \left. 
 - {\sqrt {\left(\ell + s\right)\left( \ell - s+ 1 \right) }} 
 \sum _{m=-\ell }^{\ell }  (-1)^{s-1} \Yslm{s-1}{\ell }{m}(\theta ,\phi )\Yslm{s'}{\ell }{m*}(\theta ', \phi ')
 \right\} 
	\nonumber \\ &  \hspace{-5cm} = 
	\frac{1}{2}{\sqrt {\frac{2\ell + 1}{4\pi }}}  \left\{   
	{\sqrt {\left( \ell -s \right)\left( \ell + s + 1 \right)}} \, {\rm {e}}^{-{\rm {i}}(s+1)\alpha } \Yslm{s+1}{\ell }{-s'} (\beta, \gamma ) 
	-{\sqrt {\left( \ell +s \right)\left( \ell - s + 1 \right)}} \, {\rm {e}}^{-{\rm {i}}(s-1)\alpha } \Yslm{s-1}{\ell }{-s'} (\beta, \gamma )
	\right\} .
\end{align}
\end{proof}

\begin{proof}[Derivation of \eqref{eq:addition2}]
	\begin{align}
		(-1)^{s}\sum _{m=-\ell }^{\ell } m \, \Yslm{s}{\ell }{m} (\theta , \phi ) \Yslm{s'}{\ell }{m*}(\theta ', \phi ')
		& = 
		-{\rm {i}} \, (-1)^{s}\sum _{m=-\ell }^{\ell } \left[ \frac{\partial }{\partial \phi }\Yslm{s}{\ell }{m} (\theta , \phi ) \right] 
	\Yslm{s'}{\ell }{m*}(\theta ', \phi ')
	\nonumber \\ &  \hspace{-4.5cm} = 
	\frac{(-1)^{s}}{2}\sum _{m=-\ell }^{\ell } \left[ \edth{s}  \Yslm{s}{\ell }{m}(\theta , \phi ) - \edthbar{s}  \Yslm{s}{\ell }{m}(\theta , \phi ) \right] 	\Yslm{s'}{\ell }{m*}(\theta ', \phi ') \sin \theta 
	- s  (-1)^{s} \cos \theta \sum _{m=-\ell }^{\ell } \Yslm{s}{\ell }{m} (\theta , \phi ) \Yslm{s'}{\ell }{m*}(\theta ', \phi ')
	\nonumber \\ &  \hspace{-4.5cm} = 
	\frac{(-1)^{s}}{2} \sum _{m=-\ell }^{\ell }  \left\{   
	\left[ {\sqrt {\left(\ell - s\right)\left( \ell + s+ 1 \right) }} \, \Yslm{s+1}{\ell }{m}(\theta ,\phi )
 		+ {\sqrt {\left(\ell + s\right)\left( \ell - s+ 1 \right) }} \, \Yslm{s-1}{\ell }{m}(\theta ,\phi ) \right]  \Yslm{s'}{\ell }{m*}(\theta ', \phi ')
	\right\} \sin \theta 
	\nonumber \\ & \qquad 
	- s (-1)^{s}\cos \theta \sum _{m=-\ell }^{\ell } \Yslm{s}{\ell }{m} (\theta , \phi ) \Yslm{s'}{\ell }{m*}(\theta ', \phi ')
 \nonumber \\ & \hspace{-4.5cm} = 
 \frac{1}{2} \left\{ -
 {\sqrt {\left(\ell - s\right)\left( \ell + s+ 1 \right) }} \sum _{m=-\ell }^{\ell } (-1)^{s+1}\Yslm{s+1}{\ell }{m}(\theta ,\phi )\Yslm{s'}{\ell }{m*}(\theta ', \phi ')
 \right. \nonumber \\ &  \hspace{-2cm} \left.
 -  {\sqrt {\left(\ell + s\right)\left( \ell - s+ 1 \right) }} \sum _{m=-\ell }^{\ell } (-1)^{s-1}\Yslm{s-1}{\ell }{m}(\theta ,\phi ) \Yslm{s'}{\ell }{m*}(\theta ', \phi ')
 \right\}  \sin \theta 
 \nonumber \\ &  \hspace{-1cm} 
 - s \cos \theta \sum _{m=-\ell }^{\ell }  (-1)^{s} \Yslm{s}{\ell }{m} (\theta , \phi ) \Yslm{s'}{\ell }{m*}(\theta ', \phi ')
	\nonumber \\ &  \hspace{-4.5cm} = 
	-\frac{1}{2}{\sqrt {\frac{2\ell + 1}{4\pi }}}  \left\{   
	{\sqrt {\left( \ell -s \right)\left( \ell + s + 1 \right)}} \, {\rm {e}}^{-{\rm {i}}(s+1)\alpha } \Yslm{s+1}{\ell }{-s'} (\beta, \gamma ) 
 \right. \nonumber \\ & \hspace{-2cm} \left. 
	+{\sqrt {\left( \ell +s \right)\left( \ell - s + 1 \right)}} \, {\rm {e}}^{-{\rm {i}}(s-1)\alpha } \Yslm{s-1}{\ell }{-s'} (\beta, \gamma )
	\right\}  \sin \theta  
	 - s {\sqrt {\frac{2\ell + 1}{4\pi }}} \, {\rm {e}}^{-{\rm {i}}s\alpha } \Yslm{s}{\ell }{-s'} (\beta, \gamma )\cos \theta .
\end{align}
\end{proof}

\begin{proof}[Derivation of \eqref{eq:addition3}]
\begin{align}
	(-1)^{s}\sum _{m=-\ell }^{\ell } \left[ \frac{\partial }{\partial \theta }\Yslm{s}{\ell }{m} (\theta , \phi ) \right] 
 \left[ \frac{\partial }{\partial \theta '}	\Yslm{s'}{\ell }{m*}(\theta ', \phi ')  \right]
 \nonumber \\ 
 & \hspace{-6cm} = 
 	\frac{(-1)^{s}}{4}  \sum _{m=-\ell }^{\ell } \left[ \edth{s}  \Yslm{s}{\ell }{m}(\theta , \phi ) + \edthbar{s}  \Yslm{s}{\ell }{m}(\theta , \phi ) \right] 
 	\left[ \edth{s'}'  \Yslm{s}{\ell }{m}(\theta ', \phi ') + \edthbar{s'}'  \Yslm{s}{\ell }{m}(\theta ', \phi ') \right] ^{*}
 	\nonumber \\ & \hspace{-6cm} = 
 		\frac{(-1)^{s}}{4}  \sum _{m=-\ell }^{\ell } \left[ {\sqrt {\left(\ell - s\right)\left( \ell + s+ 1 \right) }} \, \Yslm{s+1}{\ell }{m}(\theta ,\phi )
 		- {\sqrt {\left(\ell + s\right)\left( \ell - s+ 1 \right) }} \, \Yslm{s-1}{\ell }{m}(\theta ,\phi )
 			\right]
 			\nonumber \\ & \hspace{-2cm}  \times 
 \left[ {\sqrt {\left(\ell - s'\right)\left( \ell + s'+ 1 \right) }} \, \Yslm{s'+1}{\ell }{m*}(\theta ',\phi ')
 - {\sqrt {\left(\ell + s'\right)\left( \ell - s'+ 1 \right) }} \, \Yslm{s'-1}{\ell }{m*}(\theta ',\phi ')
 \right]	
 \nonumber \\ & \hspace{-6cm} = 
 -\frac{1}{4}  \sum _{m=-\ell }^{\ell } \left[  
 {\sqrt {\left(\ell - s\right)\left( \ell + s+ 1 \right) \left(\ell - s'\right)\left( \ell + s'+ 1 \right) }} \, (-1)^{s+1}\Yslm{s+1}{\ell }{m}(\theta ,\phi )\Yslm{s'+1}{\ell }{m*}(\theta ',\phi ')
 \right. \nonumber \\ & \hspace{-2cm} \left. 
 -{\sqrt {\left(\ell - s\right)\left( \ell + s+ 1 \right) \left(\ell + s'\right)\left( \ell - s'+ 1 \right) }} \, (-1)^{s+1}\Yslm{s+1}{\ell }{m}(\theta ,\phi )\Yslm{s'-1}{\ell }{m*}(\theta ',\phi ')
  \right. \nonumber \\ & \hspace{-2cm} \left. 
 -{\sqrt {\left(\ell + s\right)\left( \ell - s+ 1 \right) \left(\ell - s'\right)\left( \ell + s'+ 1 \right) }} \, (-1)^{s-1}\Yslm{s-1}{\ell }{m}(\theta ,\phi )\Yslm{s'+1}{\ell }{m*}(\theta ',\phi ')
   \right. \nonumber \\ & \hspace{-2cm} \left. 
 +{\sqrt {\left(\ell + s\right)\left( \ell - s+ 1 \right) \left(\ell + s'\right)\left( \ell -s'+ 1 \right) }} \, (-1)^{s-1}\Yslm{s-1}{\ell }{m}(\theta ,\phi )\Yslm{s'-1}{\ell }{m*}(\theta ',\phi ')
 \right] 
 \nonumber \\ & \hspace{-6cm} = 
  -\frac{1}{4}{\sqrt {\frac{2\ell + 1}{4\pi }}}   \left[
   {\sqrt {\left(\ell - s\right)\left( \ell + s+ 1 \right) \left(\ell - s'\right)\left( \ell + s'+ 1 \right) }} \, 
   {\rm {e}}^{-{\rm {i}}(s+1)\alpha }\Yslm{s+1}{\ell }{-s'-1}(\beta,\gamma )
    \right. \nonumber \\ & \hspace{-2cm} \left. 
   -{\sqrt {\left(\ell - s\right)\left( \ell + s+ 1 \right) \left(\ell + s'\right)\left( \ell - s'+ 1 \right) }} \, 
    {\rm {e}}^{-{\rm {i}}(s+1)\alpha }\Yslm{s+1}{\ell }{-s'+1}(\beta,\gamma )
   \right. \nonumber \\ & \hspace{-2cm} \left. 
   -{\sqrt {\left(\ell + s\right)\left( \ell - s+ 1 \right) \left(\ell - s'\right)\left( \ell + s'+ 1 \right) }} \, 
    {\rm {e}}^{-{\rm {i}}(s-1)\alpha }\Yslm{s-1}{\ell }{-s'-1}(\beta,\gamma )
   \right. \nonumber \\ & \hspace{-2cm} \left. 
   +{\sqrt {\left(\ell + s\right)\left( \ell - s+ 1 \right) \left(\ell + s'\right)\left( \ell -s'+ 1 \right) }} \, 
    {\rm {e}}^{-{\rm {i}}(s-1)\alpha }\Yslm{s-1}{\ell }{-s'+1}(\beta,\gamma )
  \right]	.	 
\end{align}
	\end{proof}
	
\begin{proof}[Derivation of \eqref{eq:addition4}]
	\begin{align}
		(-1)^{s}\sum _{m=-\ell }^{\ell } m^{2} \Yslm{s}{\ell }{m} (\theta , \phi )\Yslm{s'}{\ell }{m*}(\theta ', \phi ')
	 \nonumber \\ & \hspace{-4.5cm}
	 = 	(-1)^{s}\sum _{m=-\ell }^{\ell } \left[ \frac{\partial }{\partial \phi }\Yslm{s}{\ell }{m} (\theta , \phi ) \right] 
		\left[ \frac{\partial }{\partial \phi '}	\Yslm{s'}{\ell }{m*}(\theta ', \phi ')  \right]
		\nonumber \\ & \hspace{-4.5cm} = 
		(-1)^{s}\sum _{m=-\ell }^{\ell } \left\{ 	\frac{{\rm {i}}}{2} \sin \theta \left[ \edth{s} - \edthbar{s} \right] \Yslm{s}{\ell }{m}(\theta , \phi )
		- {\rm {i}}s \cos \theta \Yslm{s}{\ell }{m}(\theta , \phi ) \right\} 
		\nonumber \\ & \hspace{-3cm} \times
		\left\{ 	\frac{{\rm {i}}}{2} \sin \theta '\left[ \edth{s'}' - \edthbar{s'}' \right] \Yslm{s'}{\ell }{m}(\theta ', \phi ')
		- {\rm {i}}s' \cos \theta '\Yslm{s'}{\ell }{m}(\theta ', \phi ')  \right\}  ^{*}
		 \nonumber \\ & \hspace{-4.5cm} =
		(-1)^{s}\sum _{m=-\ell }^{\ell } 
		\bigg\{ 	\frac{{\rm {i}}}{2} \sin \theta  \left[ 
		{\sqrt {\left(\ell - s\right)\left( \ell + s+ 1 \right) }} \, \Yslm{s+1}{\ell }{m}(\theta ,\phi )
		+ {\sqrt {\left(\ell + s\right)\left( \ell - s+ 1 \right) }} \, \Yslm{s-1}{\ell }{m}(\theta ,\phi ) \right]
  \nonumber \\ & \hspace{-2cm}  
		- {\rm {i}}s \cos \theta \Yslm{s}{\ell }{m}(\theta , \phi )
		\bigg\}
			\nonumber \\ & \hspace{-4cm} \times
			\bigg\{  	-\frac{{\rm {i}}}{2} \sin \theta ' \left[  {\sqrt {\left(\ell - s'\right)\left( \ell + s'+ 1 \right) }} \, \Yslm{s'+1}{\ell }{m*}(\theta ',\phi ')
			+ {\sqrt {\left(\ell + s'\right)\left( \ell - s'+ 1 \right) }} \, \Yslm{s'-1}{\ell }{m*}(\theta ',\phi ') \right] 
			\nonumber \\ & \hspace{-2cm}  
				+ {\rm {i}}s' \cos \theta '\Yslm{s'}{\ell }{m*}(\theta ', \phi ')  \bigg\} 
				 \nonumber \\ & \hspace{-4.5cm} =
-\sum _{m=-\ell }^{\ell } \bigg\{ 
\frac{1}{4}\sin \theta \sin \theta ' \left[ 
{\sqrt {\left(\ell - s\right)\left( \ell + s+ 1 \right) \left(\ell - s'\right)\left( \ell + s'+ 1 \right) }} \, (-1)^{s+1}\Yslm{s+1}{\ell }{m}(\theta ,\phi )\Yslm{s'+1}{\ell }{m*}(\theta ',\phi ')
\right. \nonumber \\ & \hspace{-1cm} \left.  
+{\sqrt {\left(\ell - s\right)\left( \ell + s+ 1 \right) \left(\ell + s'\right)\left( \ell - s'+ 1 \right) }} \, (-1)^{s+1}\Yslm{s+1}{\ell }{m}(\theta ,\phi )\Yslm{s'-1}{\ell }{m*}(\theta ',\phi ')
 \right. \nonumber \\ & \hspace{-1cm} \left.   
+{\sqrt {\left(\ell + s\right)\left( \ell - s+ 1 \right) \left(\ell - s'\right)\left( \ell + s'+ 1 \right) }} \, (-1)^{s-1}\Yslm{s-1}{\ell }{m}(\theta ,\phi )\Yslm{s'+1}{\ell }{m*}(\theta ',\phi ')
 \right. \nonumber \\ & \hspace{-1cm} \left.  
+{\sqrt {\left(\ell + s\right)\left( \ell - s+ 1 \right) \left(\ell + s'\right)\left( \ell -s'+ 1 \right) }} \, (-1)^{s-1}\Yslm{s-1}{\ell }{m}(\theta ,\phi )\Yslm{s'-1}{\ell }{m*}(\theta ',\phi ')
\right]
\nonumber \\ & \hspace{-4cm} 
- \frac{s'}{2} \sin \theta \cos \theta ' \left[ 
{\sqrt {\left(\ell - s\right)\left( \ell + s+ 1 \right) }} \, (-1)^{s+1}\Yslm{s+1}{\ell }{m}(\theta ,\phi )
 \right. \nonumber \\  & \left. \hspace{-1cm}
+ {\sqrt {\left(\ell + s\right)\left( \ell - s+ 1 \right) }} \, (-1)^{s-1}\Yslm{s-1}{\ell }{m}(\theta ,\phi ) 
\right] \Yslm{s'}{\ell }{m*}(\theta ', \phi ')
 \nonumber \\ & \hspace{-4cm} 
+ \frac{s}{2} \cos \theta \sin \theta ' \left[
 {\sqrt {\left(\ell - s'\right)\left( \ell + s'+ 1 \right) }} \, \Yslm{s'+1}{\ell }{m*}(\theta ',\phi ')
 \right. \nonumber \\  & \left. \hspace{-1cm}
+ {\sqrt {\left(\ell + s'\right)\left( \ell - s'+ 1 \right) }} \, \Yslm{s'-1}{\ell }{m*}(\theta ',\phi ')
\right] (-1)^{s}\Yslm{s}{\ell }{m}(\theta , \phi )
 \nonumber \\ & \hspace{-4cm}  
- ss' (-1)^{s}\Yslm{s}{\ell }{m}(\theta , \phi )\Yslm{s'}{\ell }{m*}(\theta ', \phi ') \cos \theta \cos \theta '
\bigg\} 
\nonumber \\ & \hspace{-4.5cm} =
-{\sqrt {\frac {2\ell +1}{4\pi }}}\bigg\{ 
\frac{1}{4}\sin \theta \sin \theta ' \left[ 
{\sqrt {\left(\ell - s\right)\left( \ell + s+ 1 \right) \left(\ell - s'\right)\left( \ell + s'+ 1 \right) }} \,
{\rm {e}}^{-{\rm {i}}(s+1)\alpha } \Yslm{s+1}{\ell }{-s'-1}(\beta  , \gamma )
\right. \nonumber \\ & \hspace{-1cm} \left.  
+{\sqrt {\left(\ell - s\right)\left( \ell + s+ 1 \right) \left(\ell + s'\right)\left( \ell - s'+ 1 \right) }} \, 
{\rm {e}}^{-{\rm {i}}(s+1)\alpha } \Yslm{s+1}{\ell }{-s'+1}(\beta  , \gamma )
\right. \nonumber \\ & \hspace{-1cm} \left.   
+{\sqrt {\left(\ell + s\right)\left( \ell - s+ 1 \right) \left(\ell - s'\right)\left( \ell + s'+ 1 \right) }} \, 
{\rm {e}}^{-{\rm {i}}(s-1)\alpha }\Yslm{s-1}{\ell }{-s'-1}(\beta  , \gamma )
\right. \nonumber \\ & \hspace{-1cm} \left.  
+{\sqrt {\left(\ell + s\right)\left( \ell - s+ 1 \right) \left(\ell + s'\right)\left( \ell -s'+ 1 \right) }} \, 
{\rm {e}}^{-{\rm {i}}(s-1)\alpha }\Yslm{s-1}{\ell }{-s'+1}(\beta  , \gamma )
\right]
\nonumber \\ & \hspace{-4cm} 
- \frac{s'}{2} \sin \theta \cos \theta ' \left[ 
{\sqrt {\left(\ell - s\right)\left( \ell + s+ 1 \right) }} \, 
{\rm {e}}^{-{\rm {i}}(s+1)\alpha }\Yslm{s+1}{\ell }{-s'}(\beta  , \gamma )
\right. \nonumber \\ & \left. \hspace{-1cm}
+ {\sqrt {\left(\ell + s\right)\left( \ell - s+ 1 \right) }} \, 
{\rm {e}}^{-{\rm {i}}(s-1)\alpha }\Yslm{s-1}{\ell }{-s'}(\beta  , \gamma ) 
\right] 
\nonumber \\ & \hspace{-4cm} 
+ \frac{s}{2} \cos \theta \sin \theta ' \left[
{\sqrt {\left(\ell - s'\right)\left( \ell + s'+ 1 \right) }} \, 
{\rm {e}}^{-{\rm {i}}s\alpha }\Yslm{s}{\ell }{-s'-1}(\beta  , \gamma ) 
+ {\sqrt {\left(\ell + s'\right)\left( \ell - s'+ 1 \right) }} \, 
{\rm {e}}^{-{\rm {i}}s\alpha }\Yslm{s}{\ell }{-s'+1}(\beta  , \gamma ) 
\right] 
\nonumber \\ & \hspace{-4cm}  
- ss' {\rm {e}}^{-{\rm {i}}s\alpha }\Yslm{s}{\ell }{-s'}(\beta , \gamma ) \cos \theta \cos \theta '
\bigg\}  .
\end{align}
\end{proof}

\begin{proof}[Derivation of \eqref{eq:addition5}]
	\begin{align}
		(-1)^{s}\sum _{m=-\ell }^{\ell } m\left[ \frac{\partial }{\partial \theta }\Yslm{s}{\ell }{m} (\theta , \phi ) \right] 
		\Yslm{s'}{\ell }{m*}(\theta ', \phi ') &
		\nonumber \\ & \hspace{-5.5cm} =
		{\rm {i}}\, (-1)^{s}\sum _{m=-\ell }^{\ell } \left[ \frac{\partial }{\partial \theta }\Yslm{s}{\ell }{m} (\theta , \phi ) \right] 
		\left[ \frac{\partial }{\partial \phi '}	\Yslm{s'}{\ell }{m*}(\theta ', \phi ')  \right]
	\nonumber \\ & \hspace{-5.5cm} =
		-\frac{{\rm {i}}}{2} (-1)^{s}\sum _{m=-\ell }^{\ell } \left[ \edth{s}  \Yslm{s}{\ell }{m}(\theta , \phi ) + \edthbar{s}  \Yslm{s}{\ell }{m}(\theta , \phi ) \right] 
			\left\{ 	\frac{{\rm {i}}}{2} \sin \theta '\left[ \edth{s'}' - \edthbar{s'}' \right] \Yslm{s'}{\ell }{m}(\theta ', \phi ')
		- {\rm {i}}s' \cos \theta '\Yslm{s'}{\ell }{m}(\theta ', \phi ')  \right\}  ^{*}
		\nonumber \\ & \hspace{-5.5cm} =
		\frac{(-1)^{s}}{4}\sum _{m=-\ell }^{\ell } \left[
		{\sqrt {\left(\ell - s\right)\left( \ell + s+ 1 \right) }} \, \Yslm{s+1}{\ell }{m}(\theta ,\phi )
		- {\sqrt {\left(\ell + s\right)\left( \ell - s+ 1 \right) }} \, \Yslm{s-1}{\ell }{m}(\theta ,\phi )
		\right] 
		\nonumber \\ & \hspace{-4.5cm} \times 
		\bigg\{ 
		 \left[  {\sqrt {\left(\ell - s'\right)\left( \ell + s'+ 1 \right) }} \, \Yslm{s'+1}{\ell }{m*}(\theta ',\phi ')
		+ {\sqrt {\left(\ell + s'\right)\left( \ell - s'+ 1 \right) }} \, \Yslm{s'-1}{\ell }{m*}(\theta ',\phi ') \right] 
		\sin \theta '
		\nonumber \\ & \hspace{-2cm}  
		- 2s' \cos \theta '\Yslm{s'}{\ell }{m*}(\theta ', \phi ') 
		\bigg\}
		\nonumber \\ & \hspace{-5.5cm} =
		-\frac{1}{4} \sum_{m=-\ell }^{\ell } \bigg\{ 
		 \left[
		{\sqrt {\left(\ell - s\right)\left( \ell + s+ 1 \right) \left(\ell - s'\right)\left( \ell + s'+ 1 \right) }} \, (-1)^{s+1}\Yslm{s+1}{\ell }{m}(\theta ,\phi ) \, \Yslm{s'+1}{\ell }{m*}(\theta ',\phi ')
		\right. \nonumber \\ & \hspace{-2cm} \left. 
		- {\sqrt {\left(\ell + s\right)\left( \ell - s+ 1 \right) \left(\ell - s'\right)\left( \ell + s'+ 1 \right) }} \, (-1)^{s-1}\Yslm{s-1}{\ell }{m}(\theta ,\phi ) \, \Yslm{s'+1}{\ell }{m*}(\theta ',\phi ')
			\right. \nonumber \\ & \hspace{-2cm} \left.
			+ {\sqrt {\left(\ell - s\right)\left( \ell + s+ 1 \right) \left(\ell + s'\right)\left( \ell - s'+ 1 \right) }} \, (-1)^{s+1}\Yslm{s+1}{\ell }{m}(\theta ,\phi ) \, \Yslm{s'-1}{\ell }{m*}(\theta ',\phi ')
			\right. \nonumber \\ & \hspace{-2cm} \left. 
			- {\sqrt {\left(\ell + s\right)\left( \ell - s+ 1 \right) \left(\ell + s'\right)\left( \ell - s'+ 1 \right) }} \, (-1)^{s-1}\Yslm{s-1}{\ell }{m}(\theta ,\phi ) \, \Yslm{s'-1}{\ell }{m*}(\theta ',\phi ')
		\right] \sin \theta '
		\nonumber \\ & \hspace{-3.5cm} 
		-2s' \left[  {\sqrt {\left(\ell - s\right)\left( \ell + s+ 1 \right) }} \, (-1)^{s+1}\Yslm{s+1}{\ell }{m}(\theta ,\phi )
  \right. \nonumber \\  & \left. \hspace{-1cm}
		- {\sqrt {\left(\ell + s\right)\left( \ell - s+ 1 \right) }} \, (-1)^{s-1}\Yslm{s-1}{\ell }{m}(\theta ,\phi ) \right] \Yslm{s'}{\ell }{m*}(\theta ', \phi ')  \cos \theta '
		\bigg\}
		\nonumber \\ & \hspace{-5.5cm} =
		-\frac{1}{4} {\sqrt {\frac {2\ell +1}{4\pi }}} \bigg\{ 
		 \left[
		{\sqrt {\left(\ell - s\right)\left( \ell + s+ 1 \right) \left(\ell - s'\right)\left( \ell + s'+ 1 \right) }} \, 
		{\rm {e}}^{-{\rm {i}}(s+1)\alpha }\Yslm{s+1}{\ell }{-s'-1}(\beta , \gamma ) 
		\right. \nonumber \\ & \hspace{-2cm} \left. 
		- {\sqrt {\left(\ell + s\right)\left( \ell - s+ 1 \right) \left(\ell - s'\right)\left( \ell + s'+ 1 \right) }} \,
		{\rm {e}}^{-{\rm {i}}(s-1)\alpha }\Yslm{s-1}{\ell }{-s'-1}(\beta , \gamma ) 
		\right. \nonumber \\ & \hspace{-2cm} \left.
		+ {\sqrt {\left(\ell - s\right)\left( \ell + s+ 1 \right) \left(\ell + s'\right)\left( \ell - s'+ 1 \right) }} \,
		{\rm {e}}^{-{\rm {i}}(s+1)\alpha }\Yslm{s+1}{\ell }{-s'+1}(\beta , \gamma ) 
		\right. \nonumber \\ & \hspace{-2cm} \left. 
		- {\sqrt {\left(\ell + s\right)\left( \ell - s+ 1 \right) \left(\ell + s'\right)\left( \ell - s'+ 1 \right) }} \,
		{\rm {e}}^{-{\rm {i}}(s-1)\alpha }\Yslm{s-1}{\ell }{-s'+1}(\beta , \gamma ) 
		\right] \sin \theta '
		\nonumber \\ & \hspace{-3.5cm} 
		-2s' \left[  {\sqrt {\left(\ell - s\right)\left( \ell + s+ 1 \right) }} \, 
		{\rm {e}}^{-{\rm {i}}(s+1)\alpha }\Yslm{s+1}{\ell }{-s'}(\beta , \gamma ) 
  \right. \nonumber \\ & \left. \hspace{-1cm}
  - {\sqrt {\left(\ell + s\right)\left( \ell - s+ 1 \right) }} \, 
		{\rm {e}}^{-{\rm {i}}(s-1)\alpha }\Yslm{s+1}{\ell }{-s'}(\beta , \gamma ) 
		\right] \cos \theta '
		\bigg\} .
	\end{align}
	\end{proof}

\end{widetext}

\section{Conclusions}
\label{sec:conc}

In this paper we have presented some new addition theorems for \swsh\!\!, generalizing the well-known addition theorem \eqref{eq:swshaddition} \cite{Hu:1997hp,Sharma:2011fk,Michel:2020,Freeden:2022}.
These new results, like the original addition theorem, involve a sum over the azimuthal quantum number $m$, and the summand involves two \swsh\!\!, with a single derivative acting on one (or both) of these. 
Since the \swsh satisfy a second order differential equation \eqref{eq:de}, similar addition theorems involving two or more derivatives acting on a \swsh can easily be derived.
The derivation of our new results follows from a straightforward application of the raising and lowering operators $\edth{s}$ and $\edthbar{s}$ \eqref{eq:raiselower} to the original addition theorem.

Our original motivation for deriving these results was to find the sums \eqref{eq:spinsame} which we required for applications in quantum field theory on black hole space-times \cite{Alvarez,Monteverdi}.
However, our main results \eqref{eq:results} are much more general and we hope that by presenting them here they will be useful to researchers in wider application areas.

\begin{acknowledgments}
The work of A.M.~is supported by an EPSRC studentship.
The work of E.W.~is supported by STFC grant number ST/X000621/1.
No data were created or analyzed in this study.
\end{acknowledgments}

\bibliography{swsh}

\begin{thebibliography}{30}%
\makeatletter
\providecommand \@ifxundefined [1]{%
 \@ifx{#1\undefined}
}%
\providecommand \@ifnum [1]{%
 \ifnum #1\expandafter \@firstoftwo
 \else \expandafter \@secondoftwo
 \fi
}%
\providecommand \@ifx [1]{%
 \ifx #1\expandafter \@firstoftwo
 \else \expandafter \@secondoftwo
 \fi
}%
\providecommand \natexlab [1]{#1}%
\providecommand \enquote  [1]{``#1''}%
\providecommand \bibnamefont  [1]{#1}%
\providecommand \bibfnamefont [1]{#1}%
\providecommand \citenamefont [1]{#1}%
\providecommand \href@noop [0]{\@secondoftwo}%
\providecommand \href [0]{\begingroup \@sanitize@url \@href}%
\providecommand \@href[1]{\@@startlink{#1}\@@href}%
\providecommand \@@href[1]{\endgroup#1\@@endlink}%
\providecommand \@sanitize@url [0]{\catcode `\\12\catcode `\$12\catcode `\&12\catcode `\#12\catcode `\^12\catcode `\_12\catcode `\%12\relax}%
\providecommand \@@startlink[1]{}%
\providecommand \@@endlink[0]{}%
\providecommand \url  [0]{\begingroup\@sanitize@url \@url }%
\providecommand \@url [1]{\endgroup\@href {#1}{\urlprefix }}%
\providecommand \urlprefix  [0]{URL }%
\providecommand \Eprint [0]{\href }%
\providecommand \doibase [0]{https://doi.org/}%
\providecommand \selectlanguage [0]{\@gobble}%
\providecommand \bibinfo  [0]{\@secondoftwo}%
\providecommand \bibfield  [0]{\@secondoftwo}%
\providecommand \translation [1]{[#1]}%
\providecommand \BibitemOpen [0]{}%
\providecommand \bibitemStop [0]{}%
\providecommand \bibitemNoStop [0]{.\EOS\space}%
\providecommand \EOS [0]{\spacefactor3000\relax}%
\providecommand \BibitemShut  [1]{\csname bibitem#1\endcsname}%
\let\auto@bib@innerbib\@empty
\bibitem [{\citenamefont {Newman}\ and\ \citenamefont {Penrose}(1966)}]{Newman:1966ub}%
  \BibitemOpen
  \bibfield  {author} {\bibinfo {author} {\bibfnamefont {E.~T.}\ \bibnamefont {Newman}}\ and\ \bibinfo {author} {\bibfnamefont {R.}~\bibnamefont {Penrose}},\ }\bibfield  {title} {\bibinfo {title} {{Note on the Bondi-Metzner-Sachs group}},\ }\href {https://doi.org/10.1063/1.1931221} {\bibfield  {journal} {\bibinfo  {journal} {J. Math. Phys.}\ }\textbf {\bibinfo {volume} {7}},\ \bibinfo {pages} {863} (\bibinfo {year} {1966})}\BibitemShut {NoStop}%
\bibitem [{\citenamefont {Goldberg}\ \emph {et~al.}(1967)\citenamefont {Goldberg}, \citenamefont {MacFarlane}, \citenamefont {Newman}, \citenamefont {Rohrlich},\ and\ \citenamefont {Sudarshan}}]{Goldberg:1966uu}%
  \BibitemOpen
  \bibfield  {author} {\bibinfo {author} {\bibfnamefont {J.~N.}\ \bibnamefont {Goldberg}}, \bibinfo {author} {\bibfnamefont {A.~J.}\ \bibnamefont {MacFarlane}}, \bibinfo {author} {\bibfnamefont {E.~T.}\ \bibnamefont {Newman}}, \bibinfo {author} {\bibfnamefont {F.}~\bibnamefont {Rohrlich}},\ and\ \bibinfo {author} {\bibfnamefont {E.~C.~G.}\ \bibnamefont {Sudarshan}},\ }\bibfield  {title} {\bibinfo {title} {{Spin-$s$ spherical harmonics and $\eth$}},\ }\href {https://doi.org/10.1063/1.1705135} {\bibfield  {journal} {\bibinfo  {journal} {J. Math. Phys.}\ }\textbf {\bibinfo {volume} {8}},\ \bibinfo {pages} {2155} (\bibinfo {year} {1967})}\BibitemShut {NoStop}%
\bibitem [{\citenamefont {{M\"uller, C}}(1966)}]{Muller:1966}%
  \BibitemOpen
  \bibfield  {author} {\bibinfo {author} {\bibnamefont {{M\"uller, C}}},\ }\href@noop {} {\emph {\bibinfo {title} {{Spherical Harmonics}}}},\ \bibinfo {series} {Lecture Notes in Mathematics}, Vol.~\bibinfo {volume} {17}\ (\bibinfo  {publisher} {Springer-Verlag},\ \bibinfo {address} {Berlin},\ \bibinfo {year} {1966})\BibitemShut {NoStop}%
\bibitem [{\citenamefont {Torres~del Castillo}(2007)}]{TorresDelCastillo:2007}%
  \BibitemOpen
  \bibfield  {author} {\bibinfo {author} {\bibfnamefont {G.}~\bibnamefont {Torres~del Castillo}},\ }\bibfield  {title} {\bibinfo {title} {Spin-weighted spherical harmonics and their applications},\ }\href {https://www.scielo.org.mx/scielo.php?script=sci_arttext&pid=S0035-001X2007000800017} {\bibfield  {journal} {\bibinfo  {journal} {Rev. Mex. Fis.}\ }\textbf {\bibinfo {volume} {53}},\ \bibinfo {pages} {125} (\bibinfo {year} {2007})}\BibitemShut {NoStop}%
\bibitem [{\citenamefont {Hu}\ and\ \citenamefont {White}(1997)}]{Hu:1997hp}%
  \BibitemOpen
  \bibfield  {author} {\bibinfo {author} {\bibfnamefont {W.}~\bibnamefont {Hu}}\ and\ \bibinfo {author} {\bibfnamefont {M.~J.}\ \bibnamefont {White}},\ }\bibfield  {title} {\bibinfo {title} {{CMB anisotropies: total angular momentum method}},\ }\href {https://doi.org/10.1103/PhysRevD.56.596} {\bibfield  {journal} {\bibinfo  {journal} {Phys. Rev. D}\ }\textbf {\bibinfo {volume} {56}},\ \bibinfo {pages} {596} (\bibinfo {year} {1997})},\ \Eprint {https://arxiv.org/abs/astro-ph/9702170} {arXiv:astro-ph/9702170} \BibitemShut {NoStop}%
\bibitem [{\citenamefont {Lewis}\ \emph {et~al.}(2002)\citenamefont {Lewis}, \citenamefont {Challinor},\ and\ \citenamefont {Turok}}]{Lewis:2001hp}%
  \BibitemOpen
  \bibfield  {author} {\bibinfo {author} {\bibfnamefont {A.}~\bibnamefont {Lewis}}, \bibinfo {author} {\bibfnamefont {A.}~\bibnamefont {Challinor}},\ and\ \bibinfo {author} {\bibfnamefont {N.}~\bibnamefont {Turok}},\ }\bibfield  {title} {\bibinfo {title} {{Analysis of CMB polarization on an incomplete sky}},\ }\href {https://doi.org/10.1103/PhysRevD.65.023505} {\bibfield  {journal} {\bibinfo  {journal} {Phys. Rev. D}\ }\textbf {\bibinfo {volume} {65}},\ \bibinfo {pages} {023505} (\bibinfo {year} {2002})},\ \Eprint {https://arxiv.org/abs/astro-ph/0106536} {arXiv:astro-ph/0106536} \BibitemShut {NoStop}%
\bibitem [{\citenamefont {Wiaux}\ \emph {et~al.}(2006)\citenamefont {Wiaux}, \citenamefont {Jacques},\ and\ \citenamefont {Vandergheynst}}]{Wiaux:2005fp}%
  \BibitemOpen
  \bibfield  {author} {\bibinfo {author} {\bibfnamefont {Y.}~\bibnamefont {Wiaux}}, \bibinfo {author} {\bibfnamefont {L.}~\bibnamefont {Jacques}},\ and\ \bibinfo {author} {\bibfnamefont {P.}~\bibnamefont {Vandergheynst}},\ }\bibfield  {title} {\bibinfo {title} {{Fast directional correlation on the sphere with steerable filters}},\ }\href {https://doi.org/10.1086/507692} {\bibfield  {journal} {\bibinfo  {journal} {Astrophys. J.}\ }\textbf {\bibinfo {volume} {652}},\ \bibinfo {pages} {820} (\bibinfo {year} {2006})},\ \Eprint {https://arxiv.org/abs/astro-ph/0508516} {arXiv:astro-ph/0508516} \BibitemShut {NoStop}%
\bibitem [{\citenamefont {Wiaux}\ \emph {et~al.}(2007)\citenamefont {Wiaux}, \citenamefont {Jacques},\ and\ \citenamefont {Vandergheynst}}]{Wiaux:2005fm}%
  \BibitemOpen
  \bibfield  {author} {\bibinfo {author} {\bibfnamefont {Y.}~\bibnamefont {Wiaux}}, \bibinfo {author} {\bibfnamefont {L.}~\bibnamefont {Jacques}},\ and\ \bibinfo {author} {\bibfnamefont {P.}~\bibnamefont {Vandergheynst}},\ }\bibfield  {title} {\bibinfo {title} {{Fast spin +-2 spherical harmonics transforms}},\ }\href {https://doi.org/10.1016/j.jcp.2007.07.005} {\bibfield  {journal} {\bibinfo  {journal} {J. Comput. Phys.}\ }\textbf {\bibinfo {volume} {226}},\ \bibinfo {pages} {2359} (\bibinfo {year} {2007})},\ \Eprint {https://arxiv.org/abs/astro-ph/0508514} {arXiv:astro-ph/0508514} \BibitemShut {NoStop}%
\bibitem [{\citenamefont {Shiraishi}(2013)}]{Shiraishi:2012bh}%
  \BibitemOpen
  \bibfield  {author} {\bibinfo {author} {\bibfnamefont {M.}~\bibnamefont {Shiraishi}},\ }  \emph {\bibinfo {title} {Probing the Early Universe with the CMBScalar, Vector and Tensor Bispectrum}},\  \href {https://doi.org/10.1007/978-4-431-54180-6} {Springer Theses} \ (\bibinfo  {publisher} {Springer},\ \bibinfo {year} {2013})\ \Eprint {https://arxiv.org/abs/1210.2518} {arXiv:1210.2518 [astro-ph.CO]} \BibitemShut {NoStop}%
\bibitem [{\citenamefont {Seibert}(2018)}]{Seibert:2018}%
  \BibitemOpen
  \bibfield  {author} {\bibinfo {author} {\bibfnamefont {K.}~\bibnamefont {Seibert}},\ }\emph {\bibinfo {title} {Spin-weighted spherical harmonics and their application for the construction of tensor {S}lepian functions on the spherical cap}},\ \href {https://dspace.ub.uni-siegen.de/handle/ubsi/1421} {Ph.D. thesis},\ \bibinfo  {school} {Universität Siegen} (\bibinfo {year} {2018})\BibitemShut {NoStop}%
\bibitem [{\citenamefont {Thorne}(1980)}]{Thorne:1980ru}%
  \BibitemOpen
  \bibfield  {author} {\bibinfo {author} {\bibfnamefont {K.~S.}\ \bibnamefont {Thorne}},\ }\bibfield  {title} {\bibinfo {title} {{Multipole expansions of gravitational radiation}},\ }\href {https://doi.org/10.1103/RevModPhys.52.299} {\bibfield  {journal} {\bibinfo  {journal} {Rev. Mod. Phys.}\ }\textbf {\bibinfo {volume} {52}},\ \bibinfo {pages} {299} (\bibinfo {year} {1980})}\BibitemShut {NoStop}%
\bibitem [{\citenamefont {Sharma}\ and\ \citenamefont {Khanal}(2014)}]{Sharma:2011fk}%
  \BibitemOpen
  \bibfield  {author} {\bibinfo {author} {\bibfnamefont {S.~K.}\ \bibnamefont {Sharma}}\ and\ \bibinfo {author} {\bibfnamefont {U.}~\bibnamefont {Khanal}},\ }\bibfield  {title} {\bibinfo {title} {{Perturbation of FRW spacetime in NP formalism}},\ }\href {https://doi.org/10.1142/S0218271814500175} {\bibfield  {journal} {\bibinfo  {journal} {Int. J. Mod. Phys. D}\ }\textbf {\bibinfo {volume} {23}},\ \bibinfo {pages} {1450017} (\bibinfo {year} {2014})},\ \Eprint {https://arxiv.org/abs/1109.6411} {arXiv:1109.6411 [astro-ph.CO]} \BibitemShut {NoStop}%
\bibitem [{\citenamefont {Gonzalez~Ledesma}\ and\ \citenamefont {Mewes}(2020)}]{GonzalezLedesma:2020dgx}%
  \BibitemOpen
  \bibfield  {author} {\bibinfo {author} {\bibfnamefont {F.}~\bibnamefont {Gonzalez~Ledesma}}\ and\ \bibinfo {author} {\bibfnamefont {M.}~\bibnamefont {Mewes}},\ }\bibfield  {title} {\bibinfo {title} {{Spherical-harmonic tensors}},\ }\href {https://doi.org/10.1103/PhysRevResearch.2.043061} {\bibfield  {journal} {\bibinfo  {journal} {Phys. Rev. Res.}\ }\textbf {\bibinfo {volume} {2}},\ \bibinfo {pages} {043061} (\bibinfo {year} {2020})},\ \Eprint {https://arxiv.org/abs/2010.09433} {arXiv:2010.09433 [physics.class-ph]} \BibitemShut {NoStop}%
\bibitem [{\citenamefont {Spiers}\ \emph {et~al.}(2024)\citenamefont {Spiers}, \citenamefont {Pound},\ and\ \citenamefont {Wardell}}]{Spiers:2023mor}%
  \BibitemOpen
  \bibfield  {author} {\bibinfo {author} {\bibfnamefont {A.}~\bibnamefont {Spiers}}, \bibinfo {author} {\bibfnamefont {A.}~\bibnamefont {Pound}},\ and\ \bibinfo {author} {\bibfnamefont {B.}~\bibnamefont {Wardell}},\ }\bibfield  {title} {\bibinfo {title} {{Second-order perturbations of the Schwarzschild spacetime: Practical, covariant, and gauge-invariant formalisms}},\ }\href {https://doi.org/10.1103/PhysRevD.110.064030} {\bibfield  {journal} {\bibinfo  {journal} {Phys. Rev. D}\ }\textbf {\bibinfo {volume} {110}},\ \bibinfo {pages} {064030} (\bibinfo {year} {2024})},\ \Eprint {https://arxiv.org/abs/2306.17847} {arXiv:2306.17847 [gr-qc]} \BibitemShut {NoStop}%
\bibitem [{\citenamefont {Michel}\ and\ \citenamefont {Seibert}(2020)}]{Michel:2020}%
  \BibitemOpen
  \bibfield  {author} {\bibinfo {author} {\bibfnamefont {V.}~\bibnamefont {Michel}}\ and\ \bibinfo {author} {\bibfnamefont {K.}~\bibnamefont {Seibert}},\ }\bibinfo {title} {{A mathematical view on spin-weighted spherical harmonics and their applications in geodesy, in: {\it {Mathematische Geodäsie/Mathematical Geodesy. Springer Reference Naturwissenschaften}}, Freeden, W. (Ed.)}}\ (\bibinfo  {publisher} {Springer Spektrum, Berlin, Heidelberg},\ \bibinfo {year} {2020})\BibitemShut {NoStop}%
\bibitem [{\citenamefont {Freeden}\ and\ \citenamefont {Schreiner}(2022)}]{Freeden:2022}%
  \BibitemOpen
  \bibfield  {author} {\bibinfo {author} {\bibfnamefont {W.}~\bibnamefont {Freeden}}\ and\ \bibinfo {author} {\bibfnamefont {M.}~\bibnamefont {Schreiner}},\ }\bibinfo {title} {{Spin-weighted spherical harmonics, in: {\it {Spherical Functions of Mathematical Geosciences}}}}\ (\bibinfo  {publisher} {Birkhauser, Berlin, Heidelberg},\ \bibinfo {year} {2022})\BibitemShut {NoStop}%
\bibitem [{\citenamefont {Rubinstein}\ \emph {et~al.}(2015)\citenamefont {Rubinstein}, \citenamefont {Kurien},\ and\ \citenamefont {Cambon}}]{Rubinstein:2015}%
  \BibitemOpen
  \bibfield  {author} {\bibinfo {author} {\bibfnamefont {R.}~\bibnamefont {Rubinstein}}, \bibinfo {author} {\bibfnamefont {S.}~\bibnamefont {Kurien}},\ and\ \bibinfo {author} {\bibfnamefont {C.}~\bibnamefont {Cambon}},\ }\bibfield  {title} {\bibinfo {title} {Scalar and tensor spherical harmonics expansion of the velocity correlation in homogeneous anisotropic turbulence},\ }\href {https://doi.org/10.1080/14685248.2015.1051184} {\bibfield  {journal} {\bibinfo  {journal} {J.~Turbulence}\ }\textbf {\bibinfo {volume} {16}},\ \bibinfo {pages} {1058} (\bibinfo {year} {2015})}\BibitemShut {NoStop}%
\bibitem [{\citenamefont {Scanio}(1977)}]{Scanio:1977}%
  \BibitemOpen
  \bibfield  {author} {\bibinfo {author} {\bibfnamefont {J.~J.~G.}\ \bibnamefont {Scanio}},\ }\bibfield  {title} {\bibinfo {title} {Spin-weighted spherical harmonics and electromagnetic multipole expansions},\ }\href {https://doi.org/10.1119/1.10649} {\bibfield  {journal} {\bibinfo  {journal} {Am. J. Phys.}\ }\textbf {\bibinfo {volume} {45}},\ \bibinfo {pages} {173} (\bibinfo {year} {1977})}\BibitemShut {NoStop}%
\bibitem [{\citenamefont {Campbell}(1971)}]{Campbell:1971}%
  \BibitemOpen
  \bibfield  {author} {\bibinfo {author} {\bibfnamefont {W.}~\bibnamefont {Campbell}},\ }\bibfield  {title} {\bibinfo {title} {{Tensor and spinor spherical harmonics and the spin-$s$ harmonics ${}_{s}Y_{lm}(\theta, \phi )$}},\ }\href {https://doi.org/10.1063/1.1665802} {\bibfield  {journal} {\bibinfo  {journal} {J. Math. Phys.}\ }\textbf {\bibinfo {volume} {12}},\ \bibinfo {pages} {1763} (\bibinfo {year} {1971})}\BibitemShut {NoStop}%
\bibitem [{\citenamefont {Breuer}\ \emph {et~al.}(1977)\citenamefont {Breuer}, \citenamefont {Ryan},\ and\ \citenamefont {Waller}}]{Breuer:1977}%
  \BibitemOpen
  \bibfield  {author} {\bibinfo {author} {\bibfnamefont {R.}~\bibnamefont {Breuer}}, \bibinfo {author} {\bibfnamefont {M.~J.}\ \bibnamefont {Ryan}},\ and\ \bibinfo {author} {\bibfnamefont {S.}~\bibnamefont {Waller}},\ }\bibfield  {title} {\bibinfo {title} {{Some properties of spin-weighted spheroidal harmonics}},\ }\href {https://doi.org/10.1098/rspa.1977.0187} {\bibfield  {journal} {\bibinfo  {journal} {Proc. R. Soc. Lond.}\ }\textbf {\bibinfo {volume} {A 358}},\ \bibinfo {pages} {71} (\bibinfo {year} {1977})}\BibitemShut {NoStop}%
\bibitem [{\citenamefont {Dray}(1985)}]{Dray:1984gy}%
  \BibitemOpen
  \bibfield  {author} {\bibinfo {author} {\bibfnamefont {T.}~\bibnamefont {Dray}},\ }\bibfield  {title} {\bibinfo {title} {{The relationship between monopole harmonics and spin-weighted spherical harmonics}},\ }\href {https://doi.org/10.1063/1.526533} {\bibfield  {journal} {\bibinfo  {journal} {J. Math. Phys.}\ }\textbf {\bibinfo {volume} {26}},\ \bibinfo {pages} {1030} (\bibinfo {year} {1985})}\BibitemShut {NoStop}%
\bibitem [{\citenamefont {Dray}(1986)}]{Dray:1986}%
  \BibitemOpen
  \bibfield  {author} {\bibinfo {author} {\bibfnamefont {T.}~\bibnamefont {Dray}},\ }\bibfield  {title} {\bibinfo {title} {{A unified treatment of {W}igner $D$-functions, spin-weighted spherical harmonics, and monopole harmonics}},\ }\href {https://doi.org/10.1063/1.527183} {\bibfield  {journal} {\bibinfo  {journal} {J. Math. Phys.}\ }\textbf {\bibinfo {volume} {27}},\ \bibinfo {pages} {781} (\bibinfo {year} {1986})}\BibitemShut {NoStop}%
\bibitem [{\citenamefont {Boyle}(2016)}]{Boyle:2016tjj}%
  \BibitemOpen
  \bibfield  {author} {\bibinfo {author} {\bibfnamefont {M.}~\bibnamefont {Boyle}},\ }\bibfield  {title} {\bibinfo {title} {{How should spin-weighted spherical functions be defined?}},\ }\href {https://doi.org/10.1063/1.4962723} {\bibfield  {journal} {\bibinfo  {journal} {J. Math. Phys.}\ }\textbf {\bibinfo {volume} {57}},\ \bibinfo {pages} {092504} (\bibinfo {year} {2016})},\ \Eprint {https://arxiv.org/abs/1604.08140} {arXiv:1604.08140 [gr-qc]} \BibitemShut {NoStop}%
\bibitem [{\citenamefont {Lee}\ \emph {et~al.}(2023)\citenamefont {Lee}, \citenamefont {Lee},\ and\ \citenamefont {Qi}}]{Lee:2023jfi}%
  \BibitemOpen
  \bibfield  {author} {\bibinfo {author} {\bibfnamefont {B.-H.}\ \bibnamefont {Lee}}, \bibinfo {author} {\bibfnamefont {W.}~\bibnamefont {Lee}},\ and\ \bibinfo {author} {\bibfnamefont {Y.-H.}\ \bibnamefont {Qi}},\ }\bibfield  {title} {\bibinfo {title} {{Superradiance in the Kerr-Taub-NUT spacetime}},\ }\Eprint {https://arxiv.org/abs/2311.10559} {arXiv:2311.10559 [gr-qc]}  (\bibinfo {year} {2023})\BibitemShut {NoStop}%
\bibitem [{\citenamefont {Bouzas}(2011{\natexlab{a}})}]{Bouzas:2011ug}%
  \BibitemOpen
  \bibfield  {author} {\bibinfo {author} {\bibfnamefont {A.~O.}\ \bibnamefont {Bouzas}},\ }\bibfield  {title} {\bibinfo {title} {{Addition theorems for spin spherical harmonics. I Preliminaries}},\ }\href {https://doi.org/10.1088/1751-8113/44/16/165301} {\bibfield  {journal} {\bibinfo  {journal} {J. Phys. A: Math. Theor.}\ }\textbf {\bibinfo {volume} {44}},\ \bibinfo {pages} {165301} (\bibinfo {year} {2011}{\natexlab{a}})},\ \Eprint {https://arxiv.org/abs/1103.2982} {arXiv:1103.2982 [math-ph]} \BibitemShut {NoStop}%
\bibitem [{\citenamefont {Bouzas}(2011{\natexlab{b}})}]{Bouzas:2011uh}%
  \BibitemOpen
  \bibfield  {author} {\bibinfo {author} {\bibfnamefont {A.~O.}\ \bibnamefont {Bouzas}},\ }\bibfield  {title} {\bibinfo {title} {{Addition theorems for spin spherical harmonics. II Results}},\ }\href {https://doi.org/10.1088/1751-8113/44/16/165302} {\bibfield  {journal} {\bibinfo  {journal} {J. Phys. A: Math. Theor.}\ }\textbf {\bibinfo {volume} {44}},\ \bibinfo {pages} {165302} (\bibinfo {year} {2011}{\natexlab{b}})},\ \Eprint {https://arxiv.org/abs/1103.2983} {arXiv:1103.2983 [math-ph]} \BibitemShut {NoStop}%
\bibitem [{\citenamefont {Balakumar}\ \emph {et~al.}(2022)\citenamefont {Balakumar}, \citenamefont {Bernar},\ and\ \citenamefont {Winstanley}}]{Balakumar:2022yvx}%
  \BibitemOpen
  \bibfield  {author} {\bibinfo {author} {\bibfnamefont {V.}~\bibnamefont {Balakumar}}, \bibinfo {author} {\bibfnamefont {R.~P.}\ \bibnamefont {Bernar}},\ and\ \bibinfo {author} {\bibfnamefont {E.}~\bibnamefont {Winstanley}},\ }\bibfield  {title} {\bibinfo {title} {{Quantization of a charged scalar field on a charged black hole background}},\ }\href {https://doi.org/10.1103/PhysRevD.106.125013} {\bibfield  {journal} {\bibinfo  {journal} {Phys. Rev. D}\ }\textbf {\bibinfo {volume} {106}},\ \bibinfo {pages} {125013} (\bibinfo {year} {2022})},\ \Eprint {https://arxiv.org/abs/2205.14483} {arXiv:2205.14483 [hep-th]} \BibitemShut {NoStop}%
\bibitem [{\citenamefont {\'Alvarez-Dom\'\i{}nguez}\ and\ \citenamefont {Winstanley}(2024)}]{Alvarez}%
  \BibitemOpen
  \bibfield  {author} {\bibinfo {author} {\bibfnamefont {A.}~\bibnamefont {\'Alvarez-Dom\'\i{}nguez}}\ and\ \bibinfo {author} {\bibfnamefont {E.}~\bibnamefont {Winstanley}},\ }\bibfield  {title} {\bibinfo {title} {{Quantum fermion superradiance and vacuum ambiguities on charged black holes}},\ }\href@noop {} {\  (\bibinfo {year} {2024})},\ \Eprint {https://arxiv.org/abs/2411.00167} {arXiv:2411.00167 [hep-th]} \BibitemShut {NoStop}%
\bibitem [{\citenamefont {Monteverdi}\ and\ \citenamefont {Winstanley}(2024)}]{Monteverdi}%
  \BibitemOpen
  \bibfield  {author} {\bibinfo {author} {\bibfnamefont {A.}~\bibnamefont {Monteverdi}}\ and\ \bibinfo {author} {\bibfnamefont {E.}~\bibnamefont {Winstanley}},\ }\bibfield  {title} {\bibinfo {title} {{Quantum scalar field theory on equal-angular-momenta Myers-Perry-AdS black holes}},\ }\href@noop {} {\  (\bibinfo {year} {2024})},\ \Eprint {https://arxiv.org/abs/2412.02814} {arXiv:2412.02814 [hep-th]} \BibitemShut {NoStop}%
\bibitem [{\citenamefont {Khersonskii}\ \emph {et~al.}(1988)\citenamefont {Khersonskii}, \citenamefont {Moskalev},\ and\ \citenamefont {Varshalovich}}]{Khersonskii:1988krb}%
  \BibitemOpen
  \bibfield  {author} {\bibinfo {author} {\bibfnamefont {V.~K.}\ \bibnamefont {Khersonskii}}, \bibinfo {author} {\bibfnamefont {A.~N.}\ \bibnamefont {Moskalev}},\ and\ \bibinfo {author} {\bibfnamefont {D.~A.}\ \bibnamefont {Varshalovich}},\ }\href {https://doi.org/10.1142/0270} {\emph {\bibinfo {title} {{Quantum Theory Of Angular Momentum}}}}\ (\bibinfo  {publisher} {World Scientific Publishing Company},\ \bibinfo {year} {1988})\BibitemShut {NoStop}%
\end{thebibliography}%

\end{document}